\documentclass[superscriptaddress,preprint,amsmath,amssymb,footinbib]{revtex4-1}
\usepackage{graphicx}
\usepackage{bm}
\usepackage{color}
\usepackage{hyperref}
\usepackage{caption}
\usepackage{upgreek}

\newcommand{\ket}[1]{\mbox{$|#1\rangle$}}
\newcommand{\bra}[1]{\mbox{$\langle #1|$}}

\begin{document}

\title{Optical Pumping of TeH$^+$: Implications for the
Search for Varying $m_p/m_e$}

\author{Patrick R. Stollenwerk}
\affiliation{Department of Physics and Astronomy, Northwestern University, Evanston, Illinois 60208, USA}
\author{Mark G. Kokish}
\affiliation{Department of Physics and Astronomy, Northwestern University, Evanston, Illinois 60208, USA}
\author{Antonio G. S. de Oliveira-Filho}
\affiliation{Departamento de Qu\'imica, Faculdade de Filosofia, Ci\^encias e Letras de Ribeir\~ao Preto,
Universidade de S\~ao Paulo, Ribeir\~ao Preto-SP 14040-901, Brazil;}
\author{Antonio G. S. de Oliveira-Filho}
\affiliation{Departamento de Qu\'imica Fundamental, Instituto de Qu\'imica, Universidade de S\~ao Paulo, S\~ao Paulo-SP 05508-000, Brazil}
\author{Brian C. Odom}
\email{b-odom@northwestern.edu}
\affiliation{Department of Physics and Astronomy, Northwestern University, Evanston, Illinois 60208, USA}

\begin{abstract}
Molecular overtone transitions provide optical frequency transitions sensitive to variation in the proton-to-electron mass ratio ($\mu\equiv m_p/m_e$). However, robust molecular state preparation presents a challenge critical for achieving high precision. Here, we characterize infrared and optical-frequency broadband laser cooling schemes for TeH$^+$, a species with multiple electronic transitions amenable to sustained laser control.~Using rate equations to simulate laser cooling population dynamics, we~estimate the fractional sensitivity to $\mu$ attainable using TeH$^+$. We~find that laser cooling of TeH$^+$ can lead to significant improvements on current $\mu$ variation limits.
\end{abstract}
\maketitle

\section{Introduction}
The Standard Model has proven remarkably robust, but it fails to explain many known phenomena such as gravity, dark matter and dark energy.~This quandary has motivated searches for physics beyond the Standard Model, including searches for space-time evolution of the dimensionless constants. Such~an~evolution could occur over cosmic time scales~\cite{uzan2003fundamental, calmet2015cosmological} and might be related to the problem of dark energy~\cite{fritzsch2015fundamental}. Alternatively, oscillatory or transient variations over shorter time scales would be expected to arise in certain proposed models for dark matter~\cite{stadnik2015can, roberts_precision_2018}.

Molecular rotational and vibrational energies scale like $E_\textrm{h} (M/m_{e})^\beta$ where $E_\textrm{h}$ is the atomic unit of energy defined by the electronic energy scale, $M$ is the reduced mass of the molecule, $\beta = -1/2$ for vibrations and $\beta = -1$ for rotations~\cite{flambaum2007limit}. Neutrons and protons primarily derive their masses from the strong interaction such that $m_{n}\approx m_{p}\approx 3\Lambda_{QCD}$, while electrons derive their mass from the weak scale via the Higgs field vacuum expectation value~\cite{flambaum2007limit,flambaum2007enhanced}. Consequently, rotational and vibrational transitions of molecules can act as a probe into the variation of $\mu\equiv m_{p}/m_{e}$ and therefore the ratio of the strong to weak energy scales. In many models, for example models assuming Grand Unification, $\mu$ varies by a factor of 30--40 more rapidly than the fine structure constant $\alpha$. Based on these arguments, there is strong motivation for experimental searches for varying $\mu$~\cite{calmet_cosmological_nodate, flambaum_limits_2004, Jansen2014}.

Atomic hyperfine transitions also have dependence on $\mu$, but their sensitivity suffers compared with molecules because of the smaller energy interval~\cite{,Jansen2014}. However, despite orders of magnitude smaller absolute sensitivities to varying $\mu$, the simpler atomic state preparation requirements have allowed atoms to set the current best laboratory constraints. Comparison of two different hyperfine transitions and an optical atomic clock has yielded a limit of $\sim$1$\times 10^{-16}$/year~\cite{huntemann2014improved,godun_frequency_2014}.~The best experimental limit from a molecule, set by comparing a rovibrational transition in SF$_6$ to a Cs hyperfine transition, is $6\times 10^{-14}$/year~\cite{shelkovnikov2008stability}.

In TeH$^+$, a vibrational overtone transition has been identified as a potentially promising candidate for $\mu$ variation detection~\cite{kokish2017optical}. The systematic uncertainties for reasonable experimental conditions are projected at the $1\times 10^{-18}$ level or below. Furthermore, TeH$^+$ is one of a small, but growing class of molecular ions identified as having so-called diagonal Franck--Condon factors (FCFs), offering the possibility of rapid state preparation through broadband rotational cooling~\cite{nguyen2011prospects, nguyen2011challenges, stollenwerk2017electronic, kang2017ab, zhang2018ground, zhang2017spectroscopic}.

We envision the $\mu$ variation experiment being performed on a single molecular ion, using quantum logic spectroscopy (QLS)~\cite{schmidt2005spectroscopy}. Preparation of the initial spectroscopy state could be accomplished either by optical pumping~~\cite{staanum2010rotational, schneider2010all,lien_broadband_2014} or projectively~\cite{chou2017preparation}. The speed at which one can initially prepare and reset the spectroscopy state has critical implications for the statistical uncertainty that can be obtained in a measurement. Here, we evaluate realistic optical pumping state preparation timescales for TeH$^{+}$ and draw conclusions about statistical uncertainties in the search for varying $\mu$. We also discuss more generally the molecular ion qualities desirable for obtaining low statistical uncertainty and identify some molecular ion species, which can serve as benchmarks for $\mu$ variation searches.

\section{Molecular Structure}

\begin{figure}
\centering
\includegraphics[scale=1]{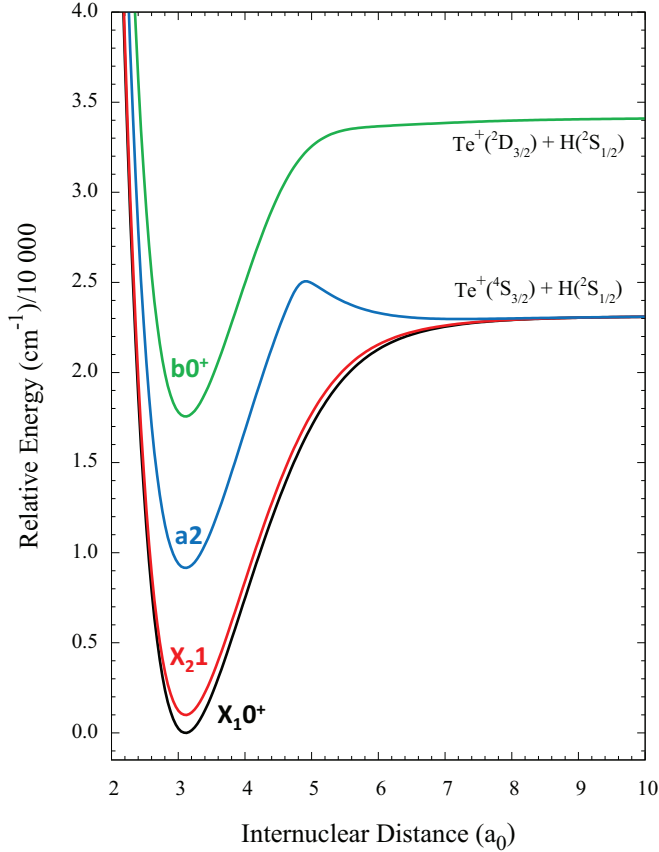}
\caption{\label{fig:pec} Low lying electronic states of TeH$^+$. Figure generated from~\cite{gonccalves2015electronic}.}
\end{figure}

\begin{figure}
\centering
\includegraphics[scale=0.55]{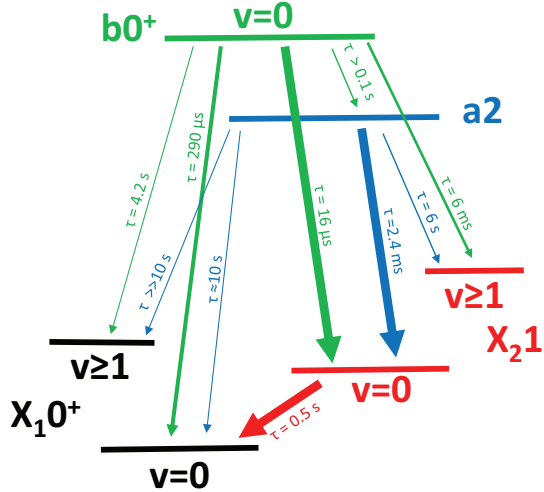}
\caption{\label{fig:Level_Diagram}
Relevant spontaneous emission channels and partial lifetimes of TeH$^+$. Line thicknesses (not to scale) represents the branching ratios of each excited state.}
\end{figure}

The four lowest lying electronic states (Figure~\ref{fig:pec}) of TeH$^+$ ($\textrm{X}_{1}0^{+}$, $\textrm{X}_{2}1$, a$_{2}$, b$0^{+}$) are well described by the Hund's case (c) basis. They are all predicted to have bond equilibrium distances within $\sim$0.1~pm of each other~\cite{gonccalves2015electronic}, implying nearly identical rotational constants and that each of the transitions will have highly diagonal FCFs. Consequently, each electronic transition will have well separated P, Q and~R branches allowing a spectrally-shaped broadband laser to selectively cover transitions that remove rotational quanta~\cite{lien_optical_2011}. Highly diagonal FCFs lead to suppressed vibrational excitation during the rotational cooling (Figure~\ref{fig:Level_Diagram}).

Since there does not exist experimental data for TeH$^+$,
we attempt to evaluate the accuracy of the TeH$^+$ multireference configuration interaction with single and double excitations and Davidson correction for higher excitations (MRCISD+Q/aV5Z) calculations~\cite{gonccalves2015electronic} by comparing theoretical~\cite{alekseyev1998spectrum} and experimental~\cite{shestakov1998lif,BEUTEL199679,yu2005infrared} investigations of the isoelectronic species antimony hydride (SbH).
Compared with the TeH$^+$ calculation, the MRCISD+Q calculation for SbH uses a smaller basis set (of quadruple zeta quality) and fewer configuration state functions and is expected to be less accurate. FCFs~depend most strongly on the difference in equilibrium bond length between electronic states, and the equilibrium bond lengths for SbH were predicted to within 3 pm of the measured values. A comparison between the predictions of the MRCISD+Q/aV5Z level of theory for the CAs molecule~\cite{DeLimaBatista2011} and experimental
measurements~\cite{Yang2011} shows that calculated bond lengths are within 1 pm of experimental values; therefore, the calculations for TeH$^+$
should be more accurate than the ones for SbH.
For optical cooling, we also rely on short lifetimes. The predicted lifetime of the b$0^{+}$ state of SbH was predicted to within a factor of two.
Other properties that have a smaller impact on cooling efficiency such as harmonic frequencies, spin-orbit splittings and electronic energies were also predicted with comparable accuracy.


Typically, multiple low-lying electronic states will complicate cooling. However, in the case of TeH$^+$, their shared diagonal FCFs open up possibilities for laser control of the internal state population using multiple broadband light sources. Diagonal transitions between the X states have energies accessible by quantum cascade lasers (QCLs), and diagonal transitions between a$_{2}$ and X and b$0^+$ and X are predicted to be in the telecom and optical bands, respectively. The lifetimes of the three low lying excited states are calculated using the potential energy curves and dipole moment functions from Gon{\c{c}}alves dos Santos et al. and LEVEL 16~\cite{le2017level} and are predicted to be 15 $\mu$s for b$0^{+}$, 2.4 ms for a2 and 460 ms for $\textrm{X}_{2}1$. Transition moments between b$0^{+}$ and a2 and between a2 and $\textrm{X}_{1}0^{+}$ will be insignificant as both are quadrupole transitions.

\subsection{Magnetic Dipole Moments}
\begin{figure}
\centering
\includegraphics[scale=1.0]{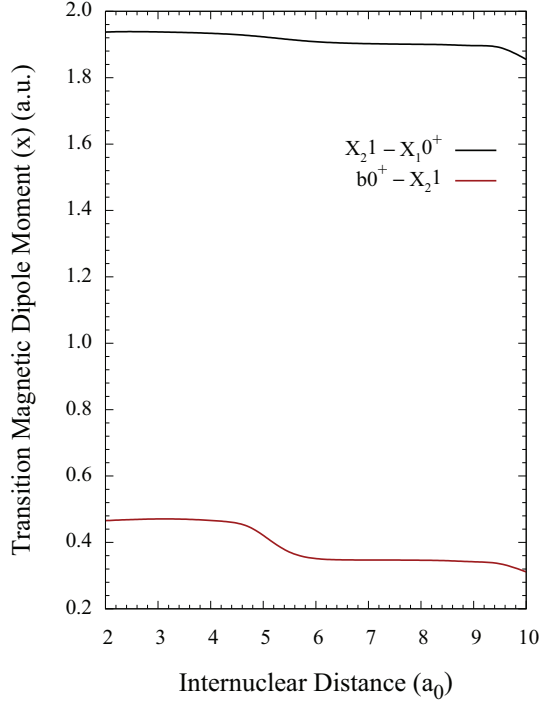}
\caption{\label{fig:magdip} Magnetic dipole transition moments of TeH$^+$.}
\end{figure}

The isoelectronic molecule SbH was observed to have significant magnetic dipole transition moments on X $\rightarrow$ b transitions~\cite{shestakov1998lif}.~Magnetic dipole transitions will connect states of the same parity, so these transitions are useful for state preparation of a single parity state, which would otherwise require an additional step to the cooling process. The TeH$^+$ magnetic dipole moments for the b$0^+-\textrm{X}_21$ ($g_s\bra{\textrm{b}0^+}S_x\ket{\textrm{X}_21}$) and $\textrm{X}_21-\textrm{X}_10^+$ ($g_s\bra{\textrm{X}_21}S_x\ket{\textrm{X}_10^+}$) transitions (Figure \ref{fig:magdip}) were computed using MOLPRO~\cite{MOLPRO} and input into LEVEL 16~\cite{le2017level} to obtain the Einstein $A$ coefficients. The~magnetic dipole spontaneous emission rates for the b$0^+-\textrm{X}_21$ and $\textrm{X}_21-\textrm{X}_10^+$ transitions are 70-times slower and five-times faster than the corresponding E1 transitions, respectively. We therefore include these M1 transitions in our simulation of the cooling dynamics.

\section{Internal State Cooling}

Generally speaking, broadband sources of light can be spectrally filtered such that frequencies driving transitions increasing vibrational or rotational energy are removed and only frequencies that drive transitions removing vibrational or rotational energy from the molecule are kept. This concept has been used previously in the cooling of rotational degrees of freedom of AlH$^+$~\cite{lien_broadband_2014} and with cooling the vibrational degrees of freedom of Cs$_2$~\cite{viteau2008optical}.~In particular, electronic transitions with diagonal FCFs undergoing spontaneous decay will tend to preserve their vibrational mode. This means that continuous pumping of rotational or vibrational energy removing transitions will efficiently populate the lowest energy rotational and vibrational states, i.e., efficient internal state cooling. In this paper, we consider variants of such a scheme on TeH$^+$. When discussing rotational cooling, we~assume a~thermal population distribution at room temperature where $\sim$99\% of the population is in $J<12$ of the ground electronic and vibrational state.

\subsection{$\textrm{X}_10^+$ $-$ $\textrm{X}_21$ Coupling}

The lifetime of the $\textrm{X}_21$ state is long compared to excited vibrational state lifetimes of $\textrm{X}_10^+$, so we do not consider rotational cooling via the $\textrm{X}_10^+ \rightarrow \textrm{X}_21$ transition. For any cooling scheme, however, $\textrm{X}_21$ will be important as it is the strongest decay channel of both b$0^+$ and a2. The addition of this laser significantly reduces the complexity of the four-level system by (1) effectively reducing $v=0$ of $\textrm{X}_10^+$ and $\textrm{X}_21$ into a single state and (2) via the relatively strong M1 transition, providing different parity coupling than in E1 transitions (Figure~\ref{fig:faster_cooling}).

Because each rotational cooling scheme must involve the population in $\textrm{X}_21$, we propose coupling $\textrm{X}_10^+$ and $\textrm{X}_21$ with a broadband laser on the Q branch. The requirements of the broadband source are simplified by the structure of $\textrm{X}_10^+$ and $\textrm{X}_21$, which has the first 12 Q branch transitions within one wavenumber of each other. The $\textrm{X}_10^+ \rightarrow \textrm{X}_21$ transition is 9.6 $\upmu$m~\cite{gonccalves2015electronic}, which allows for a single QCL to couple rotational states of the two ground vibrational states.

For cooling to proceed at the maximum rate set by upper state spontaneous emission, \mbox{the $\textrm{X}_21-\textrm{X}_10^+$} transition, whose Einstein $A$ coefficients are <~2 s$^{-1}$, must be driven at well above saturation. For a2 or b$0^{+}$ as the choice of the upper state, this requires coupling $\textrm{X}_21-\textrm{X}_10^+$ at $\sim$3 and five orders of magnitude above saturation, respectively. Given that saturation occurs with a spectral intensity of $\sim$130 $\mu$W/(mm$^2$ cm$^{-1}$), a 1~cm$^{-1}$ broad QCL with 50 mW of power focused onto the molecule is easily capable of meeting these requirements.

\subsection{Rotational Cooling on $\textrm{X}-\textrm{b}0^+$ at 600 nm}

The most rapid cooling scheme will involve the optical transitions between X and b$0^+$ as b$0^+$ has the shortest lifetime of the diagonal electronic states. We propose cooling by pumping from $\textrm{X}_21$ because the transition dipole moment of $\textrm{X}_10^+$ and b$0^+$ is expected to be an order of magnitude weaker than $\textrm{X}_21$ and b$0^+$.

\begin{figure}
\centering
\includegraphics[scale=0.5]{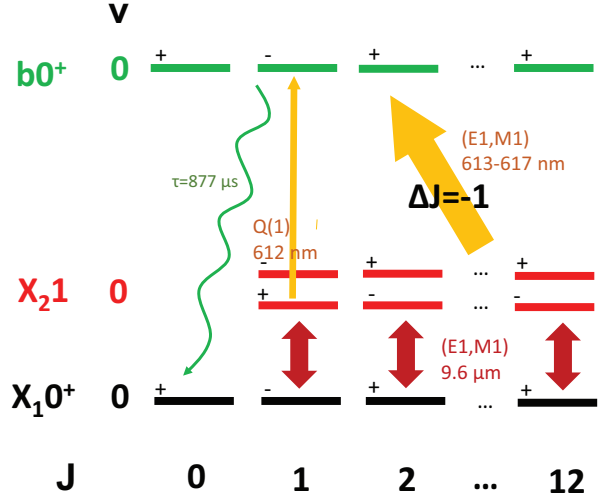}
\caption{\label{fig:faster_cooling} Rotational cooling scheme using $\textrm{X} \rightarrow \textrm{b}0^+$ at 600 nm. Straight arrows show transitions driven by lasers, with arrow width indicating laser linewidth. Both the X$_10^+$ $-$ X$_21$ coupling and the
X$_21$ $\rightarrow$ b$0^+$ lasers are capable of driving M1 transitions that preserve parity. The wavy arrow indicates the spontaneous emission channel to the dark state.}
\end{figure}

The P branch of $\textrm{X}_21$ $\rightarrow$ b$0^+$ has been predicted to span 612 nm--618 nm for $J<12$, and the spectral intensity at saturation is estimated to be $\sim$500 mW mm$^{-2}$/cm$^{-1}$ (10 W mm$^{-2}$/nm). Rapid~progress on broadband commercial lasers in this spectral region suggests that a light source capable of saturating all the required transitions might soon be available. We note that inclusion of P(1) in the coverage of the P branch with a broadband source will lead to sub-optimal cooling as decay from $\ket{\textrm{b}0^+, J=0}$ can only increase rotational energy. Exclusion of P(1), however, will limit cooling by leaving $J=1$ dark to the cooling laser. As seen in Figure~\ref{fig:faster_cooling}, this can be avoided with the addition of a CW laser tuned to $\ket{\textrm{X}_21,J=1,+}\rightarrow \ket{\textrm{b}0^+,J=1,-}$ where $J=0$ will become the only dark state. Note that the two \ket{\textrm{X}, J=1, -} states are not dark because the pump also drives M1 transitions.

\subsubsection{Vibrational Repumping}
\label{sec:repump}
Though branching from $\ket{\textrm{b}0^+,v=0}$ into $\ket{\textrm{X}_21,v=1}$ is slow, an additional CW laser and careful choice of the rotational cooling laser spectral cutoff can improve cooling time and fidelity. Because the vibrational constants of $\textrm{X}_21$ and b$0^+$ are similar, the rotational cooling laser on $\ket{\textrm{X}_21,v=0}\rightarrow \ket{\textrm{b}0^+,v=0}$ will also rotationally cool on $\ket{\textrm{X}_21,v=1}\rightarrow \ket{\textrm{b}0^+,v=1}$. The spectrum is such that the spectral cutoff can be placed between P(1) and P(2) for both vibrational states. Since the rotational cooling laser can connect states of the same parity via M1 transitions, any decays into $\ket{\textrm{X}_21,v=1}$ will therefore be pumped into $J=1$ where a CW laser can be used as a vibrational repump into $v=0$ via the P(1) transition of $\ket{\textrm{X}_21,v=1} \rightarrow \ket{\textrm{b}0^+,v=0}$ ($\sim 700$ nm). Because decays from b$0^+$ into X$_10^+$ are more than an order of magnitude less frequent
than into X$_21$, an extra laser coupling the X$_10^+$ and X$_21$ $v=1$ states is not necessary.

\subsection{Rotational Cooling on $X-a2$ at 1300 nm}

Rotational cooling with IR frequencies can be done by optical pumping through a2. The relevant $\textrm{X}_21 \rightarrow \textrm{a}2$ P branch transitions at room temperature are predicted to span $\sim$100 cm$^{-1}$ from 1340 nm--1360 nm~\cite{gonccalves2015electronic}, within the telecom O-band. The spectral intensity for saturation of these transitions is $<$~25~mW~mm$^{-2}$/cm$^{-1}$, meaning a 5~W broadband laser with a 1~mm$^2$ collimated beam area is sufficient for saturation.

Cooling via this transition will be limited by the 4--7~ms branching decay times of $\ket{a_2, J}\rightarrow \ket{\textrm{X}_21,J-1}$. A rough estimate performed by taking the number of occupied rotational states at room temperature and multiplying by the average branching time places the cooling time scale at $\sim$50 ms. Though cooling on this transition will be slow compared to cooling via the shorter lived b$0^+$ state, depending on the application, it may be advantageous given the availability of telecom technology.

\begin{figure}
\centering
\includegraphics[scale=0.5]{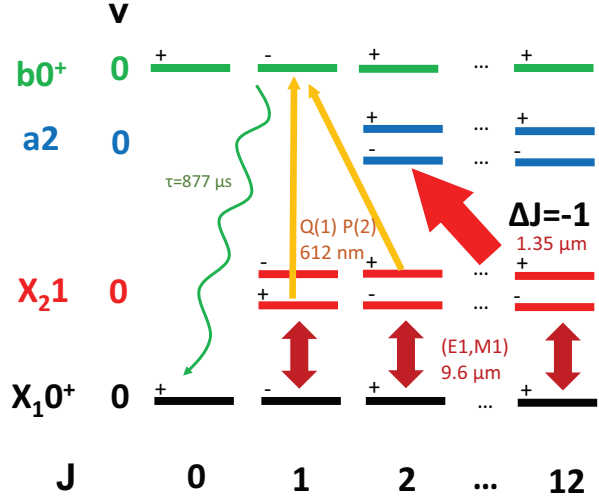}
\caption{\label{fig:slow_cooling} Rotational cooling scheme using $\textrm{X}\rightarrow \textrm{a}2$ at 1300 nm. Straight arrows show transitions driven by lasers, with arrow width indicating laser linewidth. Yellow arrows indicate CW-assist lasers. The~wavy arrow indicates the spontaneous emission channel to the dark state.}
\end{figure}

A cartoon of the transitions involved in the $\textrm{X}-\textrm{a}2$ cooling scheme(s) can be seen in Figure~\ref{fig:slow_cooling}. In~a~more careful analysis of the cooling time scale, we note that the $\textrm{X}_21 \rightarrow \textrm{a}2$ transition has no P branch transitions for $J<3$. In a cooling scheme relying on a QCL coupling the $X$ states via the Q branch and a broadband laser covering the P branch of $\textrm{X}_21 \rightarrow \textrm{a}2$, the lack of P branch transitions for $J<3$ implies rotations will cease being cooled once the population has been pumped into $J=0,1,2$. If~the broadband laser includes the Q branch of $\textrm{X}_21 \rightarrow \textrm{a}2$, then at the cost of a reduced cooling rate, the broadband laser will pump $J=2$ such that the population will transfer into $J=0,1$. Over much longer time scales (seconds) determined by the $\ket{\textrm{X}_21, J=1}\rightarrow \ket{\textrm{X}_10^+, J=0}$ branching time, the~$X$ coupling laser will pump the remaining population into $\ket{\textrm{X}_10^+, J=0}$. The fidelity of this final step will be limited by the much slower rate of blackbody redistribution.

\subsubsection{CW Assist}
With assistance from the b$0^+$ state, it is possible to avoid the rate-limiting steps that were not included in our rough estimate of the cooling time scale. In the scheme relying on pumping the P and Q branch of $\textrm{X}_21 \rightarrow \textrm{a}2$, an additional CW laser tuned to $\ket{\textrm{X}_21,J=1,+}\rightarrow \ket{\textrm{b}0^+,J=1,-}$ will more rapidly pump the population into $J$=0 than what is allowed for by the $\ket{\textrm{X}_21, J=1}\rightarrow\ket{\textrm{X}_10^+, J=0}$ branching time. Similarly, a CW laser tuned to $\ket{\textrm{X}_21, J=2,+}\rightarrow \ket{\textrm{b}0^+,J=1,-}$ can replace the role of the less efficient cooling from the $\textrm{X}_21 \rightarrow \textrm{a}2$ Q branch pumping. With the simultaneous assistance of both CW lasers, the $\textrm{X}-\textrm{a}2$ cooling scheme (Figure~\ref{fig:slow_cooling}) recovers the $\sim$50-ms time scale.

\subsection{Vibrational Cooling}
\begin{figure}
\centering
\includegraphics[scale=0.5]{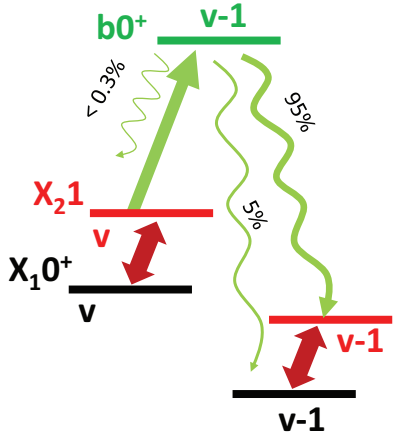}
\caption{\label{fig:v_cooling} Vibrational cooling. Straight arrows indicate transitions covered by lasers, and wavy arrows indicate spontaneous emission channels.}
\end{figure}

Depending on the choice of excited spectroscopy state in a $\mu$ variation measurement, vibrational cooling may be beneficial.~Specifically, we envision our spectroscopy states to be of the form $\ket{\textrm{X}_10^+, v''=0, J''=0}$ and $\ket{\textrm{X}_10^+, v', J'=1}$. As the excited state spontaneously decays, the rotational state population will slowly diffuse as the molecule vibrationally relaxes. For $v'=1$, decay can only leave the population in the vibrational ground state, and so, vibrational cooling is not necessary to minimize state re-preparation time. For $v'>1$ we propose active vibrational cooling by driving $\Delta v = -1$ transitions of the form $\ket{\textrm{X}_21, v} \rightarrow \ket{\textrm{b}0^+,v-1}$ (see Figure~\ref{fig:v_cooling}). Similar to the rotational cooling schemes, $\ket{\textrm{X}_10^+,v} \rightarrow \ket{\textrm{X}_21, v}$ must also be coupled, and this is accomplished via the Q branch.~However, because there is no Q branch transition for $J=0$, we must include R(0) of each vibrational level. As~the $\textrm{X}_10^+ \rightarrow \textrm{X}_21$ transitions only span $\sim$100 cm$^{-1}$ for $v=1$ to $v=7$ and the rotational spacing is large, a spectral mask blocking unwanted frequencies from a tightly-focused broadband QCL should be~sufficient.

The vibrational overtone $v=0 \rightarrow v'=8$ of TeH$^+$ has been proposed for a search for varying $\mu$~\cite{kokish2017optical}.~Cooling vibrational levels $v<8
$ requires a bandwidth of $\sim$400 cm$^{-1}$ in the 685--705-nm range. It is in principle possible to cover exclusively all of the vibrational repump transitions with a~supercontinuum laser source. Saturating the weakly-coupled off diagonal transitions, however, will~require a~spectral intensity of $> 100$ W mm$^{-2}$/ cm$^{-1}$, which is currently not commercially available. Another option is to use multiple narrower, but still broad laser sources, since each set of relevant P branch transitions span $\sim$20 cm$^{-1}$ per vibrational level.

\section{Rate Equation Simulation}

\begin{figure}
\centering
\includegraphics[scale=0.4]{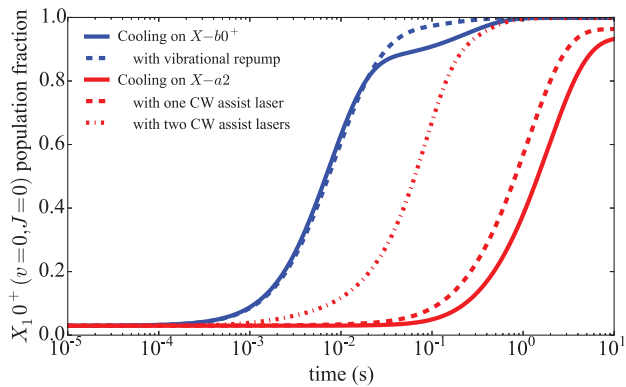}
\caption{\label{fig:cooling_times} Simulation results for the $\ket{\textrm{X}_10^+, v=0, J=0 }$ population versus cooling time, beginning from a 293 K thermal distribution.}
\end{figure}

The population distribution as a function of time for different laser cooling schemes was simulated using an Einstein $A$ and $B$ coefficient model similar to that used previously~\cite{staanum2010rotational,nguyen2011challenges,nguyen2011prospects}. The simulation includes up to 864 total states in the set of states including $\ket{\textrm{X}_10^+,v\leq 9, J \leq 15}$, $\ket{\textrm{X}_21,v\leq 7, J \leq 15, +/-}$, $\ket{\textrm{a}2,v\leq 8, J \leq 15, +/-}$ and $\ket{\textrm{b}0^+,v\leq 9, J \leq 15}$ to accurately model vibrational cascades and vibrational cooling. The model ignores hyperfine structure, and Zeeman states are treated as degenerate with their multiplicities accounted for in the Einstein coefficients. The full set of spontaneous and stimulated rates is described by an $864\times 864$ matrix that can be represented as the sum of a matrix composed exclusively of $A$ coefficients and a separate matrix using $B$ coefficients. The Einstein $A$ coefficients were calculated using LEVEL 16. The $B$ coefficients are calculated assuming a background blackbody temperature of 293 K and adding the contribution of the input spectral intensity of unpolarized laser sources at every transition wavelength. In this way, all possible incoherent coupling between states is included. Each laser source is described assuming a~Gaussian line shape with a given spectral width that is modified by a spectral mask if necessary. In~Figure \ref{fig:cooling_times}, we plot the fractional population of $\ket{\textrm{X}_10^+, v=0, J=0 }$ as a function of time under various rotational cooling schemes beginning from a 293 K Boltzmann distribution at $t = 0$.

The simplest 600~nm cooling scheme uses three lasers: (1) a 100~mW, 1~cm$^{-1}$ broad QCL coupling $\textrm{X}_10^+$ and $\textrm{X}_21$, (2) a 50~mW, 100~cm$^{-1}$ broad laser source with 3~cm$^{-1}$ spectral cutoff before the $\ket{\textrm{X}_21,v=0} \rightarrow \ket{\textrm{b}0^+,v=0}$ P(1) transition and (3) a 3-mW CW laser saturating the $\ket{\textrm{X}_21,v=0,J=1,+} \rightarrow \ket{\textrm{b}0^+,v=0,J=1,-}$ transition. The results of this combination are given by the solid blue line in Figure \ref{fig:cooling_times}. As seen in the figure, cooling in this scheme involves two primary time scales. The first time scale is the rapid cooling of rotations for the population that remains in the vibrational ground state during cooling. This results in $\sim$85\% of the population in the ground state after 25 ms. The remaining population is primarily the consequence of off-diagonal decay into $v=1$ and will slowly relax on the time scale of the $v=1$ lifetime (205 ms) such that $>$99\% is in the ground state after 1 s. The dashed blue line in Figure \ref{fig:cooling_times} is the result of our cooling simulation when we add a~vibrational repump on $\ket{\textrm{X}_21,v=1,J=1,-}\rightarrow \ket{\textrm{b}0^+,v=0, J=0,+}$. As the rotational cooling laser is still effective in the excited vibrational state, driving this lone transition is able to efficiently repump $\ket{\textrm{b}0^+,v=0}$ such that the slower time scale is no longer present.

In the 1300~nm cooling schemes, we observe a significant reduction in the cooling rate compared to the 600~nm schemes. The simplest and slowest 1300~nm scheme (solid red line Figure \ref{fig:cooling_times}) uses only two lasers: (1) the same QCL used in the optical scheme and (2) a 5~W (unfocused), 50~cm$^{-1}$ broad O-band telecom laser covering the $\textrm{X}_21\rightarrow \textrm{a}2$ P and Q branch with a 3~cm$^{-1}$ spectral cutoff before the R branch. As expected, the ground state preparation time is dominated by the slow $\ket{\textrm{X}_21, J=1}$ branching time into $\ket{\textrm{X}_10^+, J=0}$. For the two red dashed lines in Figure \ref{fig:cooling_times}, we add the CW lasers connecting $\textrm{X}_21$ to b$0^+$ to the IR cooling scheme as described previously. The longer dashed line shows the results with a laser tuned to $\ket{\textrm{X}_21,J=1,+}\rightarrow \ket{\textrm{b}0^+,J=1,-}$. The shorter dashed line shows how we recover cooling rates in line with our naive estimate by adding a laser on $\ket{\textrm{X}_21, J=2,+}\rightarrow \ket{\textrm{b}0^+,J=1,-}$ and removing the inefficient contribution of the $\textrm{X}_21\rightarrow \textrm{a}2$ Q branch.

\section{State Preparation during Spectroscopy Cycle}

In a $\mu$ variation experiment where we perform repeated measurements on the same molecule, the population distribution will be non-thermal after the spectroscopy transition is driven. We model the scenario where the spectroscopy transition is $\ket{\textrm{X}_10^+, v=0, J=0} \rightarrow \ket{\textrm{X}_10^+, v', J'=1}$, and the spectroscopy probe time is similar to the upper state lifetime $\tau$.~To optimize the cooling protocol for each choice of $v'$, we~conservatively consider that spontaneous emission from the upper state occurred at $t=0$, and the population subsequently evolved for a time $\tau$. After this simulated evolution, we~vibrationally cool and then rotationally cool before the next spectroscopy cycle. It is important to separate the two cooling stages, since simultaneous application of vibrational and rotational cooling lasers will couple separate lower states to the same excited state.~This coupling would have the unintended consequence of temporarily pumping the population into higher rovibrational states (we~note that the narrowband vibrational repumping scheme for clearing the population from $v=1$ avoids this issue by exclusively pumping to $\ket{\textrm{b}0^+, v=0, J=0}$, a state to which the rotational cooling lasers do not couple because of where we place the spectral cutoff).

The cooling time for the vibrational cooling stage is determined by minimizing the following expression:
\begin{equation}
\label{minstat}
\begin{split}
\frac{\tau+T_\textrm{VC}}{\rho_{v=0}(T_\textrm{VC})},
\end{split}
\end{equation}
where $\tau$ is the interrogation time (assumed to be equal to the excited state lifetime), $T_\textrm{VC}$ is the amount of time the vibrational cooling lasers are on and $\rho_{v=0}(t)$ is the fraction of the population in $v=0$ at time $t$. Vibrational cooling was simulated assuming broadband coverage approximately at the saturation intensity of the relevant $\ket{\textrm{X}_21, v} \rightarrow \ket{\textrm{b}0^+,v-1}$ transitions. The cooling times for the first eight excited vibrational states can be seen in Table \ref{table:vib}. In every case, the vibrational cooling lasers pumped $>99\%$ of the population into the ground vibrational state. It is noteworthy that vibrational cooling will not contribute significantly to the overall duty cycle as $T_\textrm{VC} \ll \tau$ for any choice of vibrational state.

Assuming the rotational cooling stage is applied for time $T_\textrm{RC}$, the average time for a successful experimental cycle is estimated to be:
\begin{equation}
\label{mincycle}
\begin{split}
T_\textrm{c} = 2 \frac{T_p+\tau+T_\textrm{VC}+T_\textrm{RC}}{\rho_{J=0}(T_\textrm{RC})},
\end{split}
\end{equation}
where $T_p$ is the total time necessary for state readout and hyperfine state preparation, the term $\rho_{J=0}(t)$ is the fraction of the population in $\ket{\textrm{X}_10^+, v=0, J=0 }$ at time $t$ after the start of the rotational cooling stage and the factor of two arises from needing to measure two points to estimate the offset from the line center. The optimal rotational cooling time will thus be the time that minimizes $T_c$.

\begin{table}
\caption{Properties of vibrational transitions $v=0 \rightarrow v'=n$. $T_\textrm{VC}$ is the simulated optimal cooling time for vibrational cooling. $\Omega/(2\pi)$ and $S/(2\pi)$ are in units of THz.}
\label{table:vib}
\begin{tabular}{l*{4}{c}}
\hline
$n$ & $\tau$ (ms) & $\Omega/(2 \pi)$  & $S /(2 \pi)$  & $T_\textrm{VC}$ (ms) \\
\hline
\hline
1	& 210 & 62 & 30 & 0     \\
2	& 110 & 120 & 58 & 1.0   \\
3	& 85 & 180 & 83 & 1.2   \\
4	& 70 & 230 & 110 & 1.4 \\
5 & 61 & 290 & 130 &  1.6            \\
6 & 53 & 340 & 140 &  1.7   \\
7 & 47 & 380 & 160 &  1.9            \\
8	& 41 & 430 & 170 &  2.0            \\
\hline
\hline
\end{tabular}
\end{table}

\section{\boldmath{$\mu$} Variation Measurement}

In a Ramsey measurement on a single ion, the Allan deviation is given by:
\begin{equation}
\sigma_y(T) = \frac{1}{C\Omega\, T_\textrm{R}}\sqrt{\frac{T_\textrm{c}}{2 T}},
\label{precision}
\end{equation}
where $C$ is the fringe visibility, $T_\textrm{R}$ is the Ramsey time, $T_\textrm{c}$ is the cycle time and $T$ is the total measurement time~\cite{riis2004optimum, hollberg2001optical}.~Optimal cycling occurs for $T_\textrm{c} = 2 T_\textrm{R}$ and $T_\textrm{R}$ set to about the upper state lifetime $\tau$, for~which $C \approx 0.6$~\cite{riis2004optimum}.~Laser cooling of the internal molecular state opens up the possibility for efficient state preparation, which can allow for repeated interrogation of the same molecular sample and low dead time. To evaluate the benefit of laser cooling in TeH$^+$, we estimate the statistical sensitivity to $\Delta \mu$ when using various laser cooling schemes and different vibrational overtone transitions.

The vibrational interval from $v=0$ to $v'=n$ at frequency $\Omega$ will vary in response to changing $\mu$ as described by:
\begin{equation}
\Delta \Omega = S \frac{\Delta \mu}{\mu}.
\end{equation}
Before statistics are considered, the absolute sensitivity coefficient $S = \partial{\Omega} / \partial{(\textrm{ln}\mu)}$ provides the most important figure of merit for the transition, since it expresses the shift in the measured frequency~\cite{demille2008enhanced,Zelevinsky2008}. It is also convenient to define a relative sensitivity coefficient $K_{\mu}$~\cite{Jansen2014} given by:
\begin{equation}
\frac{\Delta \Omega}{\Omega} = K_\mu \frac{\Delta \mu}{\mu}.
\end{equation}

We must also account for detrimental statistical effects of the finite upper state lifetime. Fluctuations in the frequency measurements are described by an Allan deviation $\sigma_y(T)$ for some overall measurement time $T$. The vibrational frequency measurements yield values for $\mu$ itself (albeit with a large theoretical uncertainty), and the square root of the two-sample variance in $\mu$ is:
\begin{equation}
\sigma_y^{(\mu)}(T) = \frac{\sigma_y(T)}{|K_\mu|}.
\end{equation}
Statistical uncertainty in $\mu$ variation can be related to $\sigma_y^{(\mu)}(T)$, with numerical factors depending on the details of the experimental protocol. Further details of statistical considerations for $\mu$ variation measurements using polar molecule overtone transitions are discussed in~\cite{kokish2017optical}.

\subsection{Single-Ion TeH$^+$ Measurement}
\begin{figure}
\centering
\includegraphics[scale=0.4]{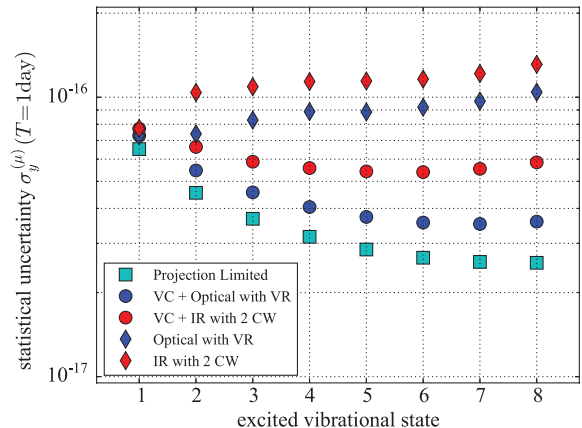}
\caption{\label{fig:foms} Simulation results for statistical uncertainty using various state preparation schemes, a~single TeH$^+$ ion and for one day of averaging.~Squares represent the projection noise-limited outcome, corresponding to instantaneous state preparation with 100\% fidelity. `Optical' indicates results for rotational cooling at 600~nm, with vibrational repump (VR) included. `IR' refers to rotational cooling at 1300~nm, with two CW-assist lasers included. Diamonds indicate results without vibrational cooling, and circles indicate results with each cycle ending with vibrational cooling (VC) followed by rotational~cooling.}
\end{figure}

In our simulated results for statistical sensitivity of a $\Delta \mu$ measurement using a single TeH$^+$ ion, the spectroscopy interval is probed using Ramsey's method, and we take $T_\textrm{R}=\tau$ and $C = 0.6$~\cite{riis2004optimum}. Results for various state preparation schemes are shown in Figure~\ref{fig:foms}. The results suggest that spectroscopy on a single TeH$^+$ ion can be used for a significantly improved search for varying $\mu$.

We find that the attainable precision is most sensitive for the larger overtone transitions. The~ultimate decision for which vibrational interval to choose for spectroscopy will depend on how much vibrational cooling laser power is available.~In the extreme case where no vibrational cooling is used, $v'=1$ is the optimal choice. At the other extreme, with enough vibrational cooling laser power to saturate all the transitions, the best simulated statistical sensitivity to $\Delta \mu$ after one day of averaging is described by $\sigma_y^{(\mu)}=3.6\times10^{-17}$. For this transition, the 600~nm cooling scheme significantly outperforms the 1300~nm cooling scheme.

\subsection{Multi-Ion Spectroscopy}

Besides searching for $\mu$ variation with a single-ion QLS measurement, an alternative approach using laser coolable polar molecules is to perform multi-ion spectroscopy. In principle, QLS can be extended to N molecular ions with only $log(N)$ overhead
in readout time and logic ions~\cite{schulte_quantum_2016}. A simpler fluorescence readout scheme is normally not possible for molecular ions, since they usually lack cycling transitions. However, for molecules that can be rapidly laser cooled, there exist quasi-cycling transitions capable of scattering enough light for fluorescence detection. Additionally, negative differential (static) polarizabilities are ubiquitous in polar molecules for transitions starting from the ground rotational state~\cite{kokish2017optical}. A negative differential polarizability allows for choosing of a magic RF trap-drive frequency such that the Stark shift and micro-motion second order Doppler shifts cancel one another~\cite{arnold2015prospects, kokish2017optical}.

\begin{figure}
\centering
\includegraphics[scale=0.4]{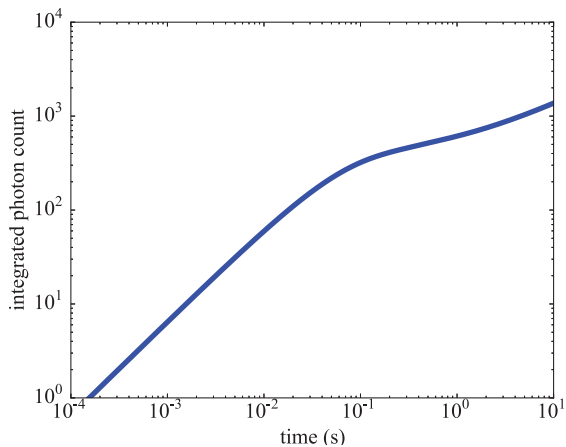}
\caption{\label{fig:photon_scatters} The total number of photon scatters from $\ket{\textrm{b}0^+, v=0, J=0 }$ as a function of time, in the two-laser fluorescence state detection scheme described in the text.}
\end{figure}

In TeH$^+$, there do exist quasi-cycling transitions amenable to state detection via fluorescence. For example, the population in \ket{\textrm{X}_10+,J=0} can be left dark, while \ket{\textrm{X}_10+,J=1} can be driven in a~quasi-cycling scheme by using one laser driving E1 and M1 coupling between $\ket{\textrm{X}_10+,J=1,-} \leftrightarrow \ket{\textrm{X}_21,J=1,\pm}$ and a second laser to couple $\ket{\textrm{X}_21,J=1,+} \leftrightarrow \ket{\textrm{b}0^+,J=0,+}$. The simulated results, using the same QCL discussed previously for the first laser and a CW laser at saturation for the second are plotted in Figure \ref{fig:photon_scatters}. On average, there will be approximately 400 spontaneously emitted photons at a~rate of $\sim$5 photons per ms before an off-diagonal $\Delta v>0$ decay occurs. In a large ensemble, the result would be a rapid decrease in the scattering rate after $\sim$80 ms.

\subsection{Homonuclear Molecule Benchmarks}

It is interesting to note that the logic of choosing the optimal overtone transition in TeH$^+$ also sets a bound on the statistical sensitivity attainable for any molecule. The strongest known chemical bond is that of CO, with $D=$ 90,000 cm$^{-1}$~\cite{luo2007comprehensive}. Molecular ion dissociation energies can approach this range; N$_2^+$ and O$_2^+$ have $D=$ 54,000 cm$^{-1}$ and $D=$ 74,000 cm$^{-1}$, respectively. For a Morse potential, the upper bound on the sensitivity $S$ is given by $D/4$, where $D$ is the dissociation energy~\cite{demille2008enhanced}. Although calculations are not generally available to describe broadening of overtone linewidths from coupling to other electronic states, the measured linewidths are expected to be limited by laser coherence. Statistical sensitivity of these species, using probe times set by currently available laser coherence, is shown in Table~\ref{table:benchmark}. Stark shifts for nonpolar species are favorably small, and other systematic uncertainties can be low, as well~\cite{kajita2016evaluation, kajita2014test, kajita2017accuracy}.

Homonuclear molecules can be loaded into the trap in the desired quantum state~\cite{tong2010sympathetic, loh2012rempi}, and~one can imagine an experimental cycle approaching zero dead time using a quantum logic protocol. Simple projective measurements within the two-level manifold can be used to reset to the lower spectroscopy state at the beginning of each cycle~\cite{chou2017preparation}. Trapped N$_2^+$ prepared in its ground rotational state lifetime has been demonstrated to have lifetimes as long as 15 minutes, limited by the collisions with background gas~\cite{tong_collisional_2011}.~After a collision changes the rotational state, a new molecule could be loaded. Alternatively, one could use a quantum logic state preparation approach that sequentially transfers the population from all possible populated states~\cite{ding2012quantum,leibfried_quantum_2012}.~In the latter approach, the problem of recovery of the pre-collision parity must also be addressed, possibly by two-photon excitation of a short-lifetime electronic transition and then cleanup of resulting vibrational excitation. Since~either state recovery approach might be time consuming, it could be preferable to operate at cryogenic temperatures to reduce the rate of collision with background gases.

Comparing the ideal zero dead time performance of TeH$^+$ and the homonuclear benchmarks in Table \ref{table:benchmark}, we find that the best TeH$^+$ statistical uncertainty is nearly two orders of magnitude larger. However, since simpler optical pumping state preparation is available for TeH$^+$, its~experimental statistical uncertainty should be less sensitive to the vacuum environment.~Furthermore, the~quasi-cycling transitions of TeH$^+$ or other polar species offers the possibility of fluorescence readout in multi-ion spectroscopy.

\begin{table}
\caption{Properties of vibrational transitions $v=0 \rightarrow v'=n$. $T_\textrm{VC}$ is the simulated optimal cooling time for vibrational cooling. $\Omega/(2\pi)$ and $S/(2\pi)$ are in units of THz.}
\label{table:vib}
\begin{tabular}{l*{4}{c}}
\hline
$n$ & $\tau$ (ms) & $\Omega/(2 \pi)$  & $S /(2 \pi)$  & $T_\textrm{VC}$ (ms) \\
\hline
\hline
1	& 210 & 62 & 30 & 0     \\
2	& 110 & 120 & 58 & 1.0   \\
3	& 85 & 180 & 83 & 1.2   \\
4	& 70 & 230 & 110 & 1.4 \\
5 & 61 & 290 & 130 &  1.6            \\
6 & 53 & 340 & 140 &  1.7   \\
7 & 47 & 380 & 160 &  1.9            \\
8	& 41 & 430 & 170 &  2.0            \\
\hline
\hline
\end{tabular}
\end{table}

\section{Conclusions}

We have identified vibrational overtone transitions in TeH$^+$ as candidates for a spectroscopic search for varying $\mu$, taking advantage of the optical pumping protocols for state preparation. Rate~equation simulations show that TeH$^+$ can be optically pumped from room temperature to the rotational ground state in $\sim$100 ms using telecom wavelengths or $\sim$10 ms using optical wavelengths. In~an~overtone spectroscopy experiment, we find that realistically achievable experimental cycle times yield a statistical uncertainty as low as $4\times10^{-17}$ for a day of averaging. This demonstrates the possibility for significant improvement on the best laboratory limit of $\sim$1 $\times 10^{-16}$/year~\cite{godun_frequency_2014, huntemann2014improved} and the current limit set by a molecule at $6\times 10^{-14}$/year~\cite{shelkovnikov2008stability}.

We primarily limited our investigation to the performance of single ion spectroscopy using quantum logic, but simulations also support the potential for fluorescence state read-out of TeH$^+$. Large Coulomb crystals of polar molecules, with state detection performed by fluorescence, could have favorably small systematic uncertainties because negative differential polarizabilities can allow for cancellation of Stark and second order Doppler shifts~\cite{arnold2015prospects}. Our analysis suggests that the possibility of searching for $\mu$ variation using multi-ion spectroscopy on laser-coolable polar species warrants further~investigation.

\vspace{6pt}

\acknowledgments{We thank Vincent Carrat for helpful discussions about laser sources. {This work was supported by ONR
 Grant No. N00014-17-1-2258, ARO
 Grant No. W911NF-14-0378 and NSF GRFP
 DGE-1324585.}




\end{document}